\documentclass[preprint,preprintnumbers,amsmath,amssymb]{revtex4}
\usepackage{graphicx}
\usepackage{epstopdf}
\usepackage{dcolumn}
\usepackage{bm}
\usepackage{color}

\begin{document}

\title{Growth and electronic structure of graphene\\ on semiconducting Ge(110)}

\author{Julia Tesch,$^1$ Elena Voloshina,$^2$ Mikhail Fonin,$^1$ and Yuriy Dedkov$^1$}

\affiliation{$^1$Fachbereich Physik, Universit\"at Konstanz, 78457 Konstanz, Germany}
\affiliation{$^2$Humboldt-Universit\"at zu Berlin, Institut f\"ur Chemie, 10099 Berlin, Germany}


\date{\today}

\begin{abstract}
The direct growth of graphene on semiconducting or insulating substrates might help to overcome main drawbacks of metal-based synthesis, like metal-atom contaminations of graphene, transfer issues, etc. Here we present the growth of graphene on $n$-doped semiconducting Ge(110) by using an atomic carbon source and the study of the structural and electronic properties of the obtained interface. We found that graphene interacts weakly with the underlying Ge(110) substrate that keeps graphene's electronic structure almost intact promoting this interface for future graphene-semiconductor applications. The effect of dopants in Ge on the electronic properties of graphene is also discussed.
\end{abstract}

\maketitle

\section{Introduction}

Presently, the main methods of the synthesis of graphene (gr), a purely 2D material consisting of carbon atoms, which can be scaled down in order to be used in further applications, are its preparation on semiconducting SiC~\cite{Virojanadara:2008,Emtsev:2009,Sprinkle:2010} or on metallic substrates~\cite{Li:2009,Bae:2010,Tao:2012tc,Dedkov:2015kp}. However, these methods have natural drawbacks like, e.\,g., the price of the high-quality SiC wafers and difficulty to control the thickness homogeneity of graphene on SiC. In case of graphene synthesis on metal substrates with the subsequent transfer onto the desired support, it was found that the level of the metal-atom contamination in the obtained graphene is not acceptable for modern microelectronics~\cite{Ambrosi:2014gl,Lupina:2015je}. These as well as other fundamental problems limit the commercialization of graphene~\cite{Peplow:2015dd,Park:2016ju} and stimulate researchers to search for the new ways of graphene synthesis.

One possibility to implement graphene in modern microelectronics processing is to perform its synthesis directly on an insulating substrate. Here one option is to use h-BN, which can be grown on the metallic substrates, like Cu, Fe, or Ni, or on semiconductors, like Ge, thus allowing a CVD synthesis of graphene, make a tunnel barrier for the carrier injection in graphene, and to avoid a metal contamination of graphene~\cite{Meng:2015jz,Kim:2015jf,Yin:2015di}.

Another approach comprises graphene synthesis directly on the semiconducting substrate. The direct growth of graphene on Si is problematic due to its carbidic phase formation at high temperatures~\cite{Hackley:2009bf,Suemitsu:2010kt,Maeda:2011bt,Hong:2012dg,ThanhTrung:2014hd}. However, the recent progress in graphene synthesis reveals the possibility to grow single- and multilayer graphene on Ge and Ge/Si substrates~\cite{Lee:2014dv,Lippert:2014fc,Rogge:2015hl,Kiraly:2015kaa,Scaparro:2016jc}. While the Ge(001) surface is the most technologically relevant one, the faceting of the underlying Ge with with the Ge($107$) facets upon graphene growth was found by means of scanning electron and tunneling microscopy (SEM and STM)~\cite{Kiraly:2015kaa,Jacobberger:2015de,Lukosius:2016ce}, that limits further technological processing of this interface. Contrary to the previous case, graphene as well as the underlying Ge surface remain flat for the Ge(110) surface, which was confirmed by low-energy electron diffraction (LEED) and STM~\cite{Lee:2014dv,Kiraly:2015kaa,Rogge:2015hl}. Despite the availability of a number of the intensive studies on the growth of graphene on Ge, the little is known about the electronic structure of this interface~\cite{Dabrowski:2016im}. In this work the \textit{ex situ} CVD grown graphene flakes on undoped Ge/Si(001) were investigated by means of micro- and nano-ARPES (angle-resolved photoelectron spectroscopy), which indicates the free-standing character of graphene maintaining the linear dispersion of the $\pi$ states in the vicinity of the Fermi level ($E_F$) and its $p$-doping with the position of the Dirac point of $E_D=0.185$\,eV above $E_F$.

Here we present a complete \textit{in situ} UHV preparation as well as structural and electronic structure study of a nearly full graphene layer epitaxially grown from an atomic carbon source on Ge(110). The presented LEED and STM results confirm the high quality of the prepared system indicating the existence of the reconstructed Ge(110) surface below graphene. Our x-ray photoelectron spectroscopy (XPS), normal-emission ARPES (NE PES), and energy-loss near-edge spectroscopy performed at the carbon $K$-edge (C $K$-edge ELNES) reveal the nearly free-standing behaviour of graphene on Ge(110). We also address the plasmon excitations in this system performing electron-energy loss spectroscopy (EELS). Our results were compared and analyzed with the available theoretical spectroscopic data for free-standing graphene and ``strongly-interacting'' gr/Ni(111) demonstrating good agreement with the former case.

\section{Experimental details}\label{Exp_Details}

Growth of graphene and all studies were performed in the surface science cluster tool (Omicron NanoTechnology; base pressure $1\times10^{-10}$\,mbar). Prior to every experiment a Ge(110) substrate (G-materials (Germany), Sb doped, resistivity $0.35\,\Omega\cdot\mathrm{cm}$) was cleaned via cycles of Ar$^+$-sputtering ($1.5$\,keV, $p(\mathrm{Ar})=1\times10^{-5}$\,mbar) and annealing ($T=870^\circ$\,C). Graphene was grown on the hot Ge(110) substrate ($T=860-870^\circ$\,C) from the atomic carbon source (Dr. Eberl MBE-Komponenten GmbH) with the filament current of $I=70$\,A and maximal pressure of $2\times10^{-9}$\,mbar during C-deposition. Cleanliness and quality of samples was controlled by LEED, STM (Omicron VT-SPM), NE PES (non-monochromatized He\,II line), and XPS (non-monochromatized Al\,$K$ line) (energy analyzer Omicron EA 125 was set either in angle-resolved or in angle-integrated mode, respectively) after every preparation step. ELNES and EELS experiments were performed in the specularly-reflected electron beam mode with angular and energy resolution of $1^\circ$ and $\approx1$\,eV, respectively. The primary electron energy is marked for every spectrum. Low-temperature (LT) STM experiments were performed in an Omicron Cryogenic STM on the gr/Ge(110) sample quickly transferred from the growth/characterization facility under N$_2$-atmosphere. Following the transfer, gr/Ge(110) was annealed in UHV at $700^\circ$\,C.

\section{Results and discussions}\label{Results_Discussions}

\begin{figure}
\center
\includegraphics[width=0.45\linewidth]{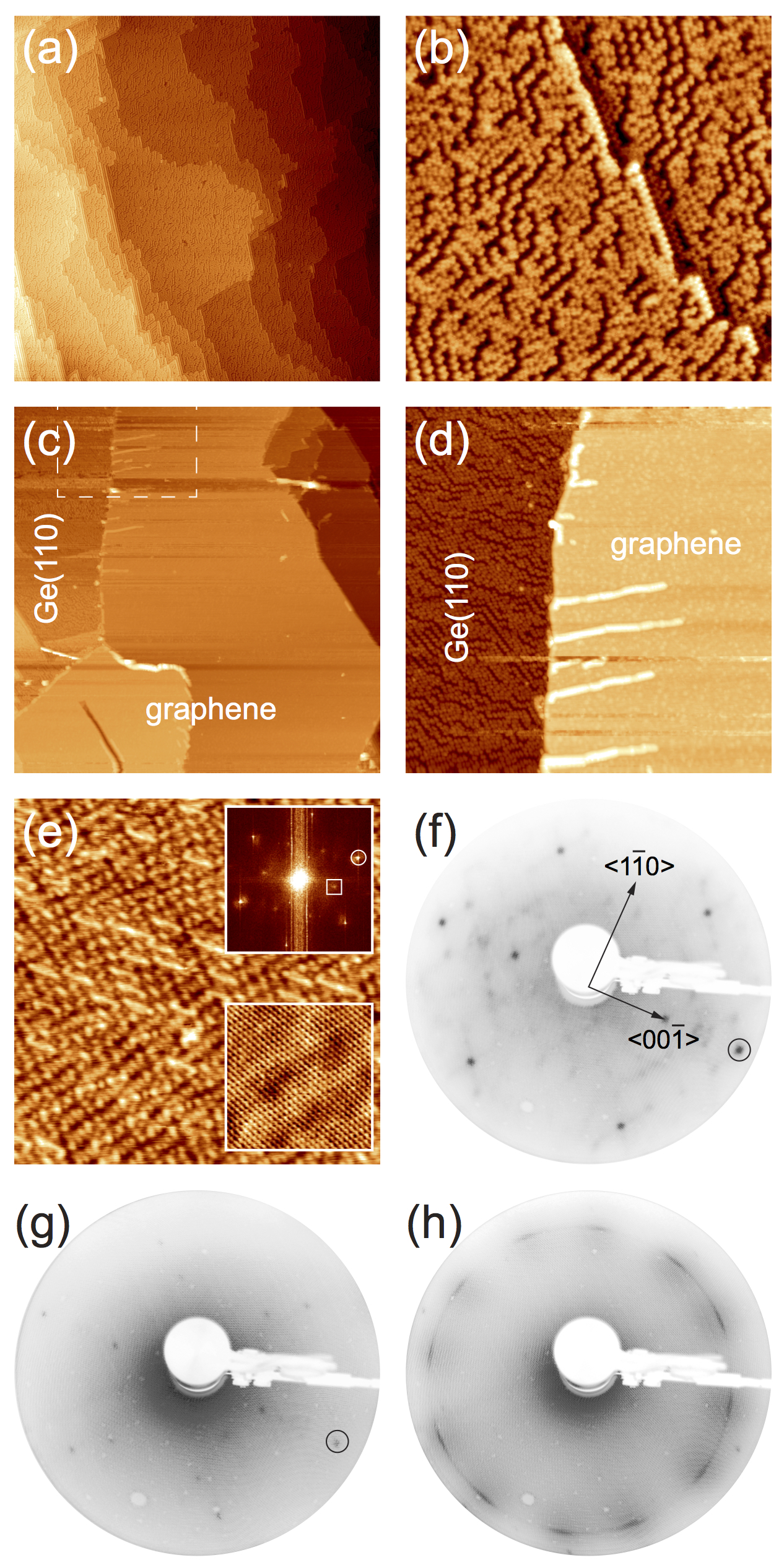}
\caption{STM and LEED characterization of Ge(110) (a,b,f) and gr/Ge(110) (c-e,g-h). Area marked by dashed rectangular in (c) is imaged with higher resolution in (d). Inset of (e) shows the corresponding FFT image of the STM data. White circle and rectangle mark the spots originating from graphene's atomic lattice and from the intervalley scattering in graphene, respectively. STM data were acquired at room temperature. Imaging parameters: (a) $500\times500\,\mathrm{nm}^2$, $U_T=+2.5$\,V, $I_T=1$\,nA, (b) $80\times80\,\mathrm{nm}^2$, $U_T=+2.5$\,V, $I_T=0.3$\,nA, (c) $400\times400\,\mathrm{nm}^2$, $U_T=+0.5$\,V, $I_T=5$\,nA, (d) $150\times150\,\mathrm{nm}^2$, $U_T=+0.5$\,V, $I_T=6$\,nA, (e) $30\times30\,\mathrm{nm}^2$, $U_T=+1.5$\,V, $I_T=0.8$\,nA (inset: $7\times7\,\mathrm{nm}^2$, $U_T=+0.02$\,V, $I_T=8$\,nA). Electron beam energy is $38$\,eV for (f,g) and $73$\,eV for (h), respectively.}
\label{Ge_grGe_STM_LEED}
\end{figure}

\begin{figure}
\center
\includegraphics[width=0.85\linewidth]{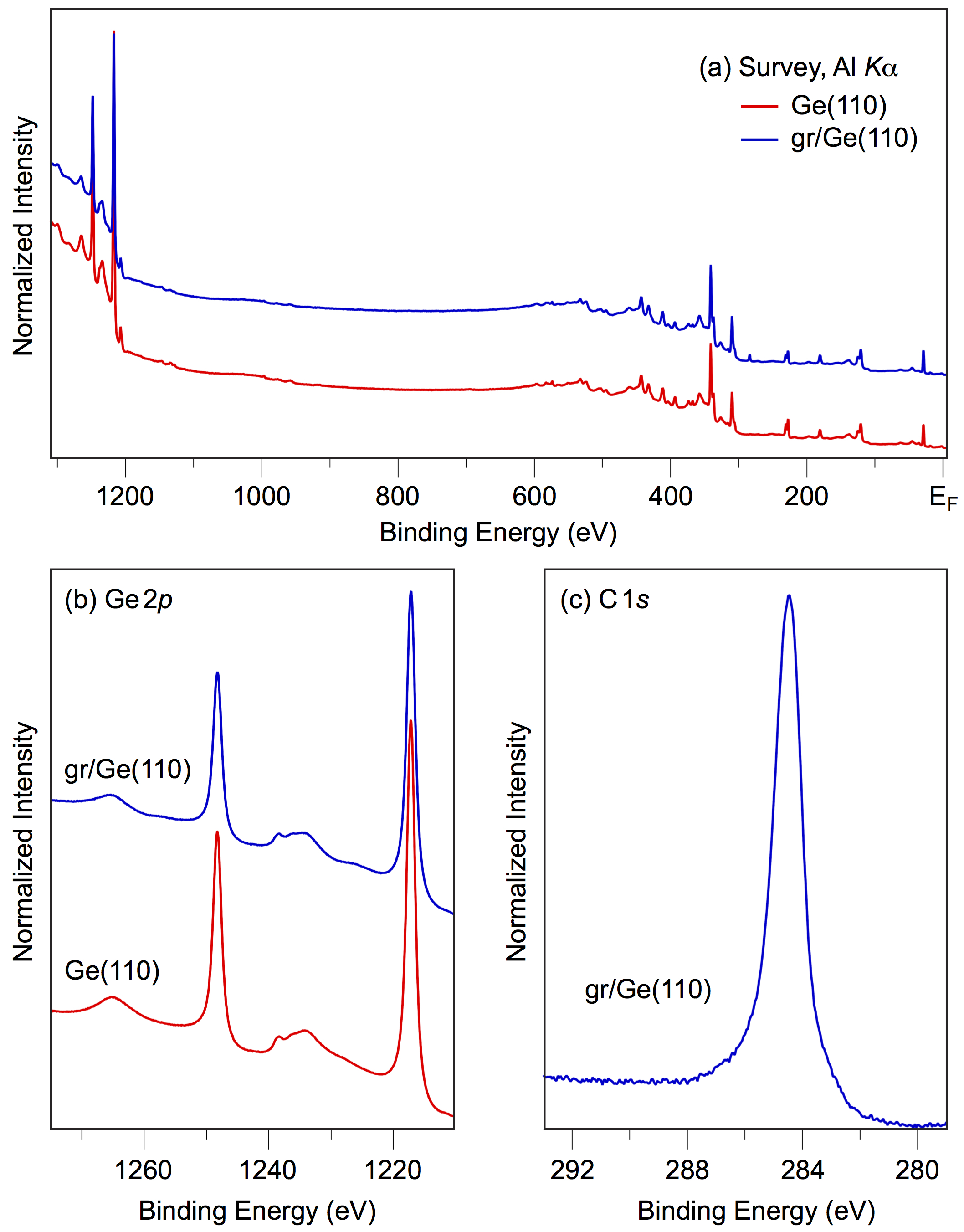}
\caption{XPS spectra of Ge(110) and gr/Ge(110): (a) surveys, (b) Ge\,$2p$, and (c) C\,$1s$.}
\label{Ge_grGe_XPS}
\end{figure}

The growth of graphene on Ge(110) was characterized by means of STM, LEED, and XPS and these results are compiled in Figs.~\ref{Ge_grGe_STM_LEED} and \ref{Ge_grGe_XPS}. The Ge(110) surface shows a large scale ordering as can be deduced from the STM [Fig.~\ref{Ge_grGe_STM_LEED}(a,b)] and LEED images [Fig.~\ref{Ge_grGe_STM_LEED}(f)]. According to previous studies this surface can be described as a faceted surface with $\{17\,15\,1\}$ facets and $c(8\times10)$ reconstruction on the steps~\cite{Olshanetsky:1977da,Ichikawa:2004kt,Ichikawa:2004cw,Mullet:2014bd}. Deposition of carbon on Ge(110) at $T=870^\circ$\,C and subsequent cooling of the sample to room temperature lifts the previously observed reconstruction, however, producing an ordered underlying Ge surface as can be seen from the respective STM and LEED images [Fig.~\ref{Ge_grGe_STM_LEED}(c-e,g)]. The prepared graphene layer forms two types of domains rotated by $30^\circ$ with respect to each other as seen from LEED and demonstrates clear honeycomb $sp^2$ structure on the Ge(110) surface [Fig.~\ref{Ge_grGe_STM_LEED}(c-e,h)]. Our results on the observation of two graphene domains are consistent with the previously reported data for the CVD grown graphene on Ge(110)~\cite{Kiraly:2015kaa}. The observed alignment of the graphene lattices of two domains is different by $\approx15^\circ$ compared to the one observed for the single-domain graphene growth in Ref.~\cite{Rogge:2015hl}. Similar to the results presented in this work, our growth methods rule out the influence of hydrogen on the alignment of graphene on Ge(110); however, further structural studies are required. Our atomically resolved STM images demonstrate clear signatures of quasiparticle scattering in the graphene layer due to imperfections in graphene as well as due to the presence of the scattering centres at the interface (segregated dopants, see discussion below). The interference of the scattering waves of the carriers in graphene leads to the formation of the corresponding $(\sqrt{3}\times\sqrt{3})R30^\circ$ structure with respect to the graphene atomic-related structure in the 2D Fast-Fourier-Transformation (FFT) map. The spots of these structures are marked in the inset of Fig.~\ref{Ge_grGe_STM_LEED}(e) by white rectangle and circle, respectively. This $(\sqrt{3}\times\sqrt{3})R30^\circ$ structure in the FFT map is assigned to the so-called intervalley scattering between adjacent cones at $K$ and $K'$ points of the graphene-derived Brillouin zone.

Formation of the uniform graphene $sp^2$ structure is also confirmed by XPS data (Fig.~\ref{Ge_grGe_XPS}). High-temperature deposition of graphene on Ge(110) leads only to the damping of the Ge\,$2p$ XPS signal [Fig.~\ref{Ge_grGe_XPS}(a,b)] without indication of the formation of the Ge-C bonds as can be concluded from the analysis of the Ge-related XPS peaks. Our data reveal a single C\,$1s$ peak for gr/Ge(110) with a small shoulder at the low binding energies (due to the possible bonds between carbon atoms and dopant atoms segregated at the interface) that confirms the homogeneity of the prepared gr/Ge(110) system.

\begin{figure}
\center
\includegraphics[width=0.85\linewidth]{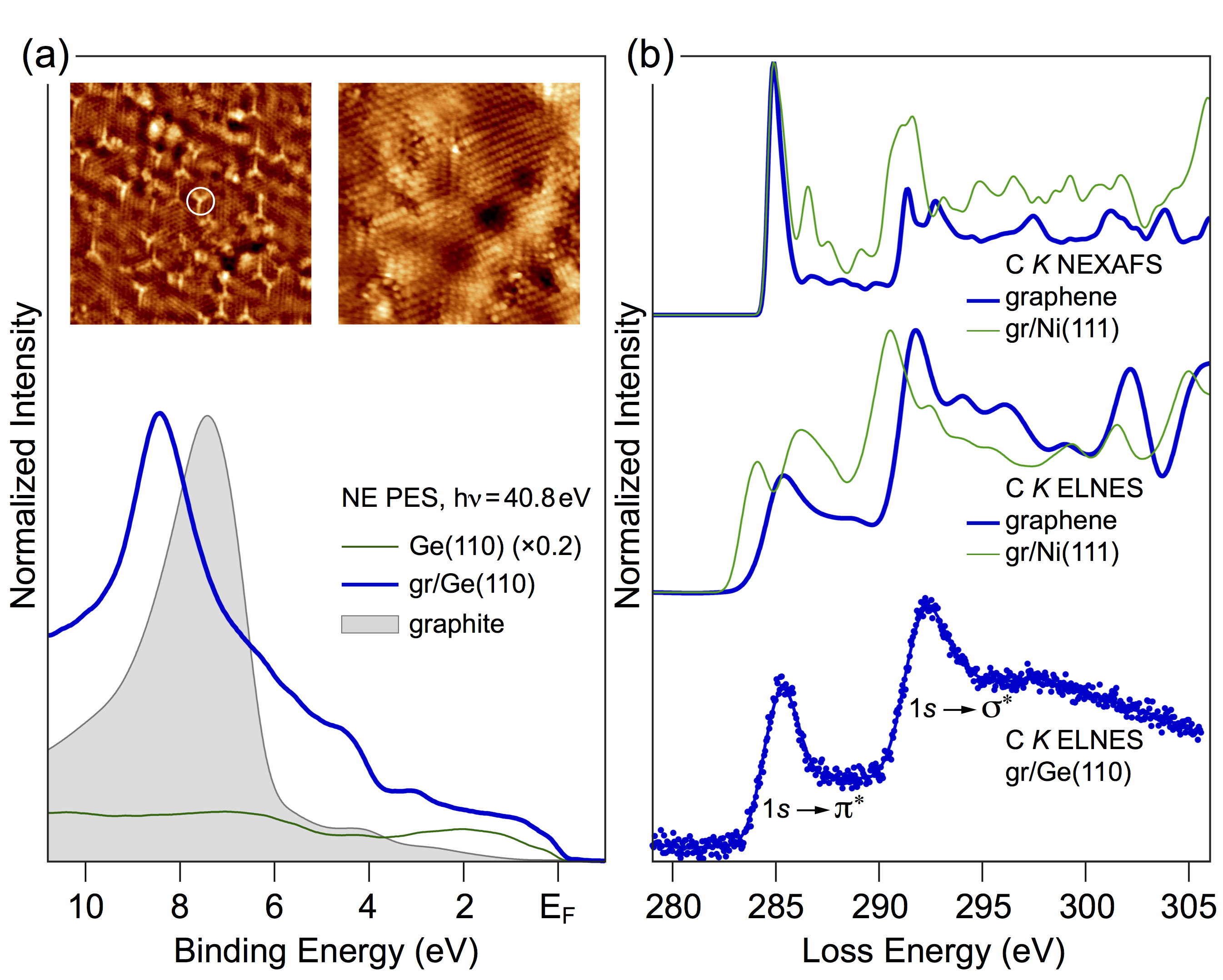}
\caption{(a) NE PES spectra of Ge(110) (intensity is scaled down by factor 5) and gr/Ge(110). Spectrum of graphite crystal is shown as a shaded area for comparison. Inset shows LT STM images of gr/Ge(110), where scattering features due to dopant atoms at the interface are clearly resolved. Imaging parameters: (left) $20\times20\,\mathrm{nm}^2$, $U_T=+1.0$\,V, $I_T=0.2$\,nA, (right) $10\times10\,\mathrm{nm}^2$, $U_T=+0.5$\,V, $I_T=0.9$\,nA. (b) Experimental and theoretical C $K$-edge ELNES and NEXAFS spectra of gr/Ge(110), graphene, and gr/Ni(111).}
\label{grGe_UPS_clEELS}
\end{figure}

The electronic structure of the grown graphene layer on Ge(110) was investigated by NE PES for the occupied valence band states below $E_F$ and by C $K$-edge ELNES for the unoccupied states above $E_F$ and these results are presented in Fig.~\ref{grGe_UPS_clEELS}(a,b), respectively. From the comparison of the PES spectrum of gr/Ge(110) and the one for the graphite single crystal we can conclude that in the former case the graphene-derived $\pi$ and $\sigma$ states are shifted to the higher binding energies by $\approx1$\,eV and $\approx0.5$\,eV, respectively. This shift indicates that in the present study the graphene layer is $n$-doped, which is opposite to the result presented in Ref.~\cite{Dabrowski:2016im}, where small $p$-doping of graphene was observed with the position of the Dirac point of $E_D=0.185$\,eV above $E_F$. This difference can be assigned to the different types of substrates used in the experiments: $n$-doped (Sb) Ge(110) in the present study and an undoped Ge-epilayer on Si(001) in Ref.~\cite{Dabrowski:2016im}. Here we can conclude that the cleaning procedure of Ge(110) (cycles of the Ar$^+$-sputtering and annealing) as well as the high temperature used during graphene growth can lead to the segregation of Sb atoms at the gr/Ge(110) interface, thus influencing the doping of the formed graphene layer. This is confirmed by our LT STM ($T=24$\,K) data of gr/Ge(110) which are presented as upper insets of Fig.~\ref{grGe_UPS_clEELS}(a), where one can clearly see the characteristic STM-signatures of such interface-trapped dopant atoms (one of them is circled). Although the first ARPES data on the Sb-atoms adsorption on graphene/SiC pointed towards the possible $p$-doping of graphene~\cite{Gierz:2008}, the recent theoretical works on the Sb intercalation in gr/SiC reveal the $n$-doping of graphene~\cite{Hsu:2013iu}. A similar effect of the $n$-doping of the free-standing graphene upon Sb adsorption was also observed in experiment~\cite{Khalil:2014dt}.

The unoccupied electronic states of graphene on Ge(110) were probed by the C $K$-edge ELNES spectroscopy, which can be considered as a simplified version of the near-edge x-ray absorption spectroscopy (NEXAFS). Here we used an electron beam of energy $E_p=700$\,eV and detected the signal originating from the energy losses due to the excitation of electrons from the C\,$1s$ core level of carbon atoms in graphene onto unoccupied states above $E_F$. Similarly to NEXAFS, this method is element-specific, i.\,e. the intensity of the loss-signal is proportional to the atom-projected partial density of unoccupied states of the element in the system, the core-level of which is involved in the process. In our case we will observe two structures, which can be assigned to the $1s\rightarrow\pi^*$ and $1s\rightarrow\sigma^*$ transitions and the respective density of states above $E_F$~\cite{Suenaga:2001bx,Titantah:2005cz,Mkhoyan:2009ff,Cupolillo:2016bt}.

The C $K$-edge ELNES spectrum of gr/Ge(110), collected in the specular-reflected electron-beam geometry, is shown in the lower part of [Fig.~\ref{grGe_UPS_clEELS}(b)] and compared with the theoretical ELNES (middle part)~\cite{Bertoni:2004} and NEXAFS (upper part)~\cite{Voloshina:2013cw} spectra of graphene and the gr/Ni(111) system. [All theoretical spectra were shifted by the same energy value in order to have the first peak, corresponding to the $1s\rightarrow\pi^*$ transition in the theoretical ELNES spectra, energetically coincide with the same peak in the experimental spectrum. The double-peak structure of the $1s\rightarrow\sigma^*$ transition in the NEXAFS spectrum is due to excitonic effects.] One can see that there is a very good agreement between experimental ELNES spectrum of gr/Ge(110) and theoretical ELNES spectrum for free-standing graphene (lower and middle parts): (i) both $1s\rightarrow\pi^*$ and $1s\rightarrow\sigma^*$ transitions exhibit a single peak at the respective threshold, that can be taken as a signature of the weak interaction between graphene and the Ge(110) surface, (ii) the energy splitting between two transitions in the experimental spectrum is almost identical to the one deduced from the theoretically calculated ELNES spectrum. As was shown in Refs.~\cite{Bertoni:2004,Weser:2010,Rusz:2010,Voloshina:2013cw} the value of this splitting as well as the modification of the shape of the $1s\rightarrow\pi^*$ transition can be taken as an indication for the $sp^2-sp^3$ rehybridization of carbon atoms, which can appear due to the graphene contact with substrate or due to the adsorption of different species on top of graphene~\cite{Dedkov:2015kp}. Such example of the spectral shape modifications of the ELNES and NEXAFS spectra for the \textit{strongly} interacting gr/Ni(111) interface is shown in Fig.~\ref{grGe_UPS_clEELS}(b). As was shown, besides the strong $n$-doping of graphene on Ni, there is a strong intermixing of the valence band states of graphene and Ni, leading to the strong modification of the energy distribution of the partial density of states of both elements. All discussed effects are clearly visible in ELNES as well as in the NEXAFS spectra, due to the similarity of the electron excitation processes. 

\begin{figure}
\center
\includegraphics[width=0.6\linewidth]{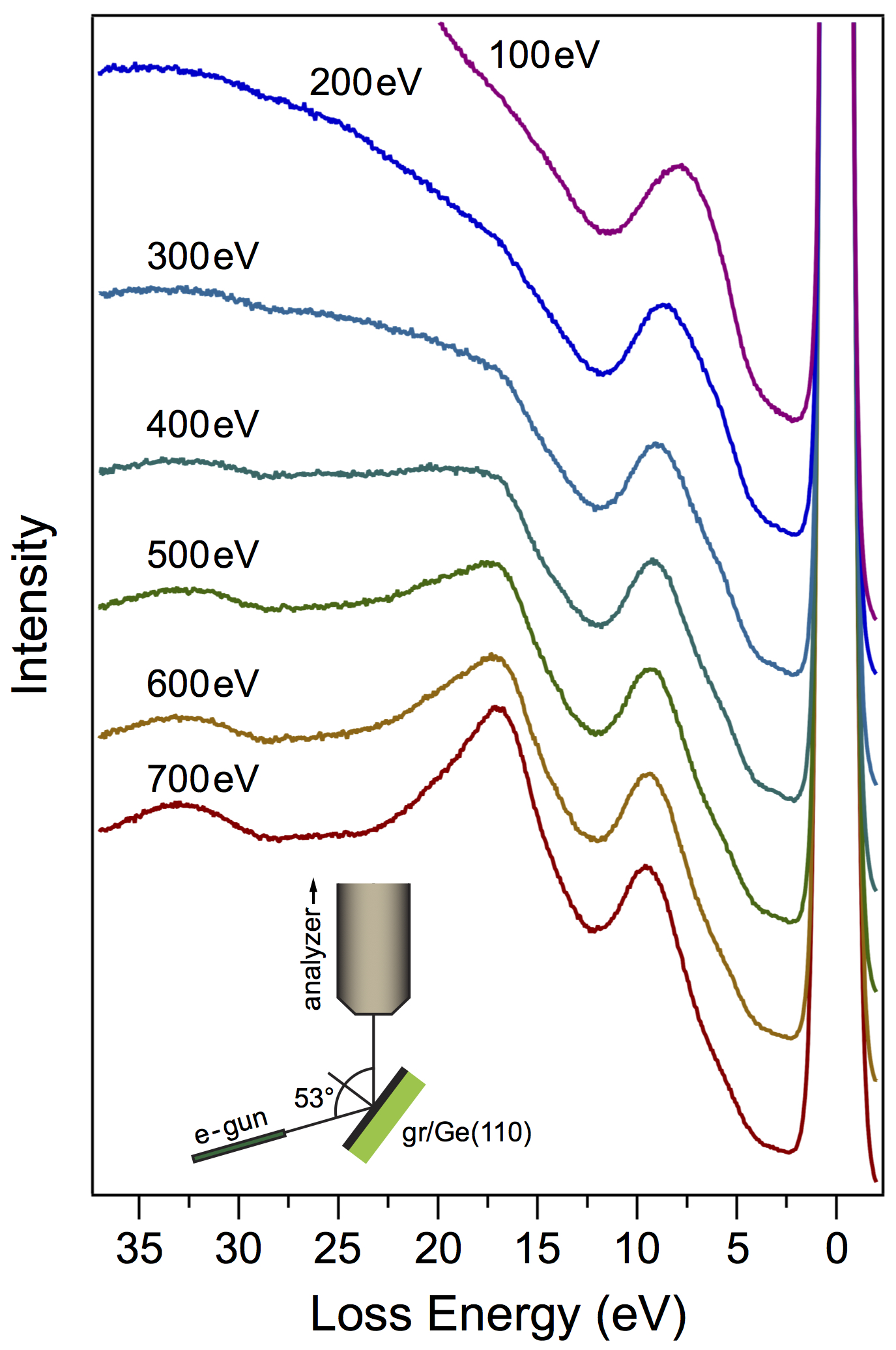}
\caption{EELS spectra of gr/Ge(110) obtained with different primary beams. The energy of the electron beam is marked for every spectra. Lower inset presents the geometry used in the EELS/ELNES experiments.}
\label{grGe_EELS_plasmons}
\end{figure}

In our EELS experiments on gr/Ge(110) we also address the plasmon excitations in the system. Figure~\ref{grGe_EELS_plasmons} shows the energy-loss spectra for this system measured as a function of the primary electron beam energy (marked for every spectrum) and presented in the energy range around the elastic peak (zero energy-loss energy). These spectra reveal a series of peaks ($\approx17$\,eV, $\approx33$\,eV), which can be clearly assigned to the bulk Ge plasmons, whereas the peak at $\approx9.5$\,eV and low energy shoulders can be assigned to the surface-related transitions of Ge(110)~\cite{Ludeke:1974iu,Ludeke:1976fa,Zhang:1993hc,Pasquali:1997hp}.

Variation of the primary beam energy allows to change the surface sensitivity of EELS as can be seen from Fig.~\ref{grGe_EELS_plasmons}. This leads to the increase of the graphene-related signal in the EELS spectra as the energy of the electron beam is decreased, which manifests itself as an increase of the intensity in the energy range of $3.5-6.5$\,eV as well as in the increase of the overall background for the energies above $15$\,eV. The first feature is assigned to the so-called $\pi$ plasmon~\cite{Lu:2009hb,Politano:2014es,Politano:2015ho}, the energy of which is determined as $6.33\pm0.25$\,eV by a curve fitting procedure. The second observation is connected to the increase of the intensity of the $\pi+\sigma$ plasmon as well as the increase of the background of the low energy inelastically scattered electrons. The exact position of the $\pi+\sigma$ plasmon cannot be extracted from these data.

\section{Conclusions}

In conclusion, we demonstrate the growth of a high-quality graphene layer on Ge(110) by evaporation of atomic carbon on the hot Ge surface. Our STM and LEED data confirm the honeycomb $sp^2$ structure of the graphene layer. From the analysis of the electronic structure of the graphene layer by means of PES and ELNES we conclude the nearly free-standing character of graphene which was found to be $n$-doped due to the segregation of Sb dopant atoms at the gr/Ge interface during sample preparation routines. Such effect of the substrate-dopant segregation at the graphene-semiconductor interface can be used for a controllable doping of graphene that might influence its electron- and spin-transport properties.

\section*{Acknowledgement}

We thank the German Research Foundation (DFG) for financial support within the Priority Programme 1459 ``Graphene''.


\clearpage

\clearpage

\clearpage

\clearpage

\end{document}